\documentclass[pra,aps,showpacs,twocolumn]{revtex4}

\usepackage{epsfig}
\usepackage{amsmath}
\usepackage{amssymb}
\begin{document}

\title{Limitations on building single-photon--resolution detection devices} 

\author{Pieter Kok} \email{Pieter.Kok@jpl.nasa.gov}

\affiliation{Quantum Computing Technologies Group, Jet Propulsion 
Laboratory, California Institute of Technology \\ Mail Stop 126-347, 
4800 Oak Grove Drive, Pasadena, California 91109-8099}

\date{\today}

\begin{abstract}
 Single-photon resolution (SPR) detectors can tell the difference between
 incoming wave packets of $n$ and $n+1$ photons. Such devices are especially
 important for linear optical quantum computing with projective measurements.
 However, in this paper I show that it is impossible to construct a
 photodetector with single-photon resolution when we are restricted to
 single-photon sources, linear optical elements and projective measurements
 with standard (non-photon-number discriminating) photodetectors. These devices
 include SPR detectors that sometimes fail to distinguish one- and two-photon
 inputs, but at the same time indicate this failure.
\end{abstract}


\maketitle

\noindent
Linear optical quantum computing and quantum communication gained
considerable momentum with the work of Knill, Laflamme and Milburn
\cite{klm01}. They showed that linear optics and projective
measurements can efficiently implement quantum computations without
the use of Kerr nonlinearities, which are typically very weak. 
Two key ingredients of
linear optical quantum computing (LOQC) are the faithful creation of
basic quantum states, such as single-photon states, and the reliable
detection of optical output states.  

Recently, several groups have demonstrated single-photon sources in
quantum-dot microcavities and magneto-optical traps
\cite{santori02,kuhn02,pelton02}. The importance of these experiments
for linear optical quantum computing can hardly be overstated, even
though the road to high-visibility interference between independent
sources is still long and arduous. Furthermore, Hockney, Dowling and I
assessed the quality of single-photon sources by defining the {\em
  suitability} of a source with respect to a given application
\cite{hockney03}. In this paper, I turn my attention to the other
essential component of LOQC: a photodetector with single-photon
resolution. 
 
Most detectors that are currently used in optical quantum communication
and computation experiments cannot tell the difference between one or more 
photons. Single-photon resolution (SPR) detectors are devices that can
distinguish between wave packets containing $n$ and $n+1$ photons
\cite{kok01,haderka03}. They are important because the LOQC research program
relies heavily on projective measurements, which in turn involve
photon-number measurements \cite{lee02,kok02}. We therefore need a way 
to efficiently 
distinguish between different photon number states. Very often, the 
output of an optical gate is significantly different when postselected 
on a single-photon or a two-photon detection outcome. Perhaps the most
dramatic example of this is the teleportation experiment by Bouwmeester 
{\em et al}.\ \cite{bouwmeester97}, in which the lack of single-photon
resolution reduces the non-postselected fidelity of the teleported
output state to a value lower than the clasical limit
\cite{braunstein98,kok00}. As a consequence, postselection was
needed. Recently, Pan {\em et al}.\ modified the pair-production rates
of the two down-converters in this experiment, such that the
non-postselected output fidelity surpasses the classical limit
\cite{zeilinger03}.  

However, it may not always be possible to modify the quantum gate in
such a way that non-SPR detectors can be used, and we really would
like to have cheap, reliable, and efficient SPR detection devices. But the 
current experimental detectors with single-photon resolution are expensive 
and operate at low temperatures \cite{takeuchi99}. In this paper, I will
therefore investigate whether we can build an SPR detector with
single-photon sources, linear optics and ordinary photodetection
without single-photon resolution. It turns out that such a device is
impossible.

Before I proceed with the proof of this statement, I will first simplify the 
task of the SPR detection device. For many applications it is good enough to
distinguish between zero-, one- and two-photon states, rather than
the general $n$- and $(n+1)$-photon states. Consequently, I will 
consider only devices that can tell the difference between one- and
two-photon states. After all, showing that such a device does not
exist immediately excludes the possibility of a general SPR detection
device. Also, the SPR device does not have to work with 100\%
efficiency to be useful (ordinary photodetectors do not, and they are
very useful). I will only require that a failure to distinguish
between a one- and two-photon input state results in an unambiguous
detector signature indicating this failure. This excludes the so-called 
{\em detector cascade} \cite{kok01,haderka03}, which splits the incoming 
mode into many outgoing modes to render the probability of finding two 
photons in the same detector arbitrarily small. However, we can never be 
certain that two photons didn't enter the same output mode, and as a 
consequence, there is no unambiguous detector signature that indicates 
failure. Here, I will prove that we cannot make an SPR device with a finite 
number of optical modes that {\em always} signals a possible failure 
unambiguously.

\begin{figure}[t]
  \begin{center}
	\epsfig{file=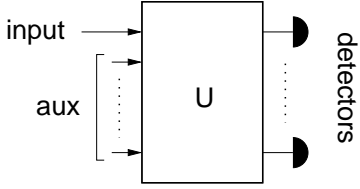, height=2.5cm}
  \end{center}
  \caption{The general setup for the single-photon resolution
     detection device with linear optics and projective
     measurements. The aim of this detector is to distinguish between
     one and two photons in the input mode. The auxiliary modes are
     occupied by photon-number states, and the detectors click only
     when one or more photons are present.} 
  \label{fig1}
\end{figure}

So far, I have been rather vague about the resources that I include for 
building an SPR device, so I will now specify them in more detail. A
schematic drawing of the prospective SPR device is shown in Fig.~\ref{fig1}, 
and consists of $M$ input modes, one of which is the mode we want to detect. 
Without loss of generality, we can choose this to be mode 1. The other 
modes contain $N-1$ auxiliary photons. So for a single-photon input state, 
a total of $N$ photons are distributed over $M$ modes according to 
$\vec{n} = (n_2,\ldots,n_M)$, with $n_i \in \Bbb{N}$ and $\sum_i n_i = N-1$. 

The $M$ modes are then transformed by a unitary transformation $U$ into 
$M$ output modes. Every annihilation operator $\hat{a}_i$ corresponding 
to input mode $a_i$ then becomes a sum of annihilation operators 
$\hat{b}_k$ that correspond to the output modes $b_k$:
\begin{equation}\label{bog}
 \hat{a}_i \quad \rightarrow\quad\sum_{k=1}^M U_{ik}\, \hat{b}_k\; .
\end{equation}
Here, we write the transformation in terms of the annihilation operators, 
but we could also have used the creation operators $\hat{a}_i^{\dagger}$ 
and $\hat{b}^{\dagger}_k$. The commutation relations are the usual:
\begin{eqnarray}
 [\hat{a}_i,\hat{a}_j^{\dagger}] &=& [\hat{b}_i,\hat{b}_j^{\dagger}] = 
 \delta_{ij} \cr [\hat{a}_i,\hat{a}_j] &=& [\hat{b}_i,\hat{b}_j] = 
 [\hat{a}_i^{\dagger},\hat{a}_j^{\dagger}] =
 [\hat{b}_i^{\dagger},\hat{b}_j^{\dagger}] = 0 \; ,
\end{eqnarray}
and all the other commutators vanish. The fact that Eq.~(\ref{bog}) does 
not mix creation and annihilation operators is due to the exclusion 
of squeezing in the $M$-port $U$. 

In principle, we can include feed-forward techniques in the transformation 
$U$. An initial unitary transformation $U_0$ is then followed by a detection 
of a subset of the outgoing modes. Based on the outcome of this detection, a 
second unitary transformation $U_1$ is applied to the remaining undetected 
modes. Again, we detect a subset of the outgoing modes. This procedure repeats 
itself until all the modes are detected. However, this technique improves only 
the efficiency, and does not increase the versatility of the device, because 
for every successful set of detector signatures and unitary transformations 
$U_0, ~U_1,\ldots$, we can postpone the intermediate detection events to the 
very end \cite{kok00b}. 

All the outgoing modes are detected with photodetectors that lack 
single-photon resolution. This amounts to a Projection Operator Valued 
Measure (POVM) of finding a click $\hat{E}_1$ and finding no click 
$\hat{E}_0$ \cite{kok00}:
\begin{eqnarray}
 \hat{E}_0 &=& \sum_{n=0}^{\infty} (1-\eta^2)^n |n\rangle\langle n| \cr
 \hat{E}_1 &=& \sum_{n=0}^{\infty} [1-(1-\eta^2)^n] |n\rangle\langle n|\; ,
\end{eqnarray}
where $\eta$ is the quantum efficiency of the detector (that is, 
every incoming photon triggers the detector with probability $\eta^2$). 
Since the objective of the proof is to show that SPR detectors cannot be 
constructed with linear optics and non-photon-resolution detectors, I 
can choose $\eta\rightarrow 1$. This corresponds to perfectly efficient 
detectors. If we can't do it with these, we certainly can't do it with less 
sensitive detectors. The POVM thus becomes
\begin{equation}
 \hat{E}_0 = |0\rangle\langle 0| \quad\text{and}\quad
 \hat{E}_1 = \sum_{n=1}^{\infty} |n\rangle\langle n|\; .
\end{equation}
The detector signature of the SPR device that should differentiate between 
one and two input photons is a string $\vec{d} = (d_1,\ldots,d_M)$, 
where $d_i \in\{ \text{`click'}, \text{`no click'} \}$. Any detector 
signature therefore belongs to one and only one of three sets: the set of 
signatures that indicate a single photon in the input mode, the set of 
signatures that indicate two photons, and the set of signatures that 
indicate a detector failure. This leads to the following criteria:

There are two ways to distinguish one- and two-photon input states:
(1) for at least one detector, the output {\em can be} a click when
one photon enters the device, but does never click when two photons
enter the device; (2) for at least one detector, the output {\em can
  be} a click when two photons enter the device, but does never click 
when only one photon enters the device. For convenience, let's call
these two methods type I and type II SPR detectors, respectively. I
will now prove that it is impossible to distinguish between one- and
two-photon input states using either type.  

In order to develop a feel for the mechanism of the proof, I will first
consider the simplest nontrivial interferometric setup. Suppose that
$a_1$ is the input mode that we want to detect, and that there is only
a single auxiliary photon in mode $a_2$. The two input modes are
transformed according to   
\begin{equation}
  \hat{a}_1 \rightarrow \alpha\,\hat{b}_1 + \beta\,\hat{b}_2
  \quad\text{and}\quad \hat{a}_2 \rightarrow \gamma\,\hat{b}_1 +
  \delta\,\hat{b}_2 \; .
\end{equation}
A single-photon input with one auxiliary photon then yields the
polynomial 
\begin{equation}\label{2single}
  \hat{a}_1 \hat{a}_2 ~\rightarrow~ \alpha\gamma\, \hat{b}_1^2 +
  (\alpha\delta + \beta\gamma)\, \hat{b}_1 \hat{b}_2 + \beta\delta\,
  \hat{b}_2^2 \; ,
\end{equation}
whereas two input photons yield
\begin{eqnarray}\label{2double}
  \hat{a}^2_1 \hat{a}_2 &\rightarrow& \alpha^2\gamma\, \hat{b}_1^3 +  
  (2\alpha\beta\gamma + \alpha^2\delta)\, \hat{b}_1^2 \hat{b}_2 \cr 
  && \quad + (2\alpha\beta\delta + \beta^2\gamma)\, \hat{b}_1
  \hat{b}_2^2 + \beta^2\delta\, \hat{b}_2^3 \; .
\end{eqnarray}
If the detection device is {\em not} to fire at a two-photon input,
all the terms involving $\hat{b}_1$ must have zero coefficients (we can
choose $\hat{b}_1$ without loss of generality). We start with the
leading term $\hat{b}_1^3$.

For the polynomial in Eq.~(\ref{2double}) to have no $\hat{b}_1^3$
contribution, we need to put $\alpha^2\gamma = 0$, which implies
either $\alpha=0$ or $\gamma=0$. Secondly, we require that
$2\alpha\beta\delta + \beta^2\gamma = 0$. If $\alpha$ and $\gamma$ are
not zero simultaneously, we also arrive at the conclusion that $\beta
= 0$ or $\delta = 0$. With these restrictions, it is no longer
possible to choose $(\alpha, \beta, \gamma, \delta)$ such that the
single-photon input of Eq.~(\ref{2single}) retains a non-zero
amplitude in the mode $b_1$.  

The general mechanism of the theorem is therefore as follows: 
consider first an SPR device of type I. By
requiring zero amplitudes in mode $b_1$ in the case of a two-photon
input state (plus the auxiliary photons), we force the coefficients of
the Bogoliubov transformation to zero. Putting coefficients to zero
will in turn force us to put other coefficients to zero. This
generates a contradiction with the non-zero $b_1$-amplitude in the
case of a single-photon input state. I then repeat the same argument for 
type II SPR devices.

First, I will consider state discrimination with a type I SPR
detector, that is, for at least one detector the output can be a
click when one photon enters the device, and does never click when two
photons enter the device. I again place this
detector in output mode $b_1$. Suppose further that we have $N-1$
auxiliary photons in modes $a_2, \ldots, a_M$, distributed according
to $\vec{n} = (n_2,\ldots,n_M)$. The transformation of the input modes
is then given by 
\begin{equation}\label{trans}
  \hat{a}_j ~\rightarrow~ \sum_{k=1}^M \alpha_{jk} \hat{b}_k\; ,
\end{equation}
where $M$ is the total number of optical input modes, and 
the matrix elements $\alpha_{jk}$ constitute a unitary matrix. 

Using Eq.~(\ref{trans}), the input of the SPR detection device in the presence of a two-photon
input state is transformed into the following polynomial:
\begin{equation}
  \hat{a}_1^2 \prod_{m=2}^N \hat{a}_m^{n_m} \rightarrow \left(
  \sum_{k=1}^M \alpha_{1k} \hat{b}_k \right)^2 \prod_{m=2}^N \left(
  \sum_{j=1}^M \alpha_{mj} \hat{b}_j \right)^{\!\! n_m} . 
\end{equation}
Again, we need to suppress the amplitudes of all the terms that
involve a factor $\hat{b}_1^k$. We start with the leading term
$\hat{b}_1^{N+1}$. There is only one term that leads to all the
photons ending up in mode $b_1$, and its coefficient must be forced to
zero:
\begin{equation}
  \alpha_{11}^2 \alpha_{21}^{n_2} \cdots \alpha_{M1}^{n_M} = 0\; .
\end{equation}
This means that $\alpha_{p1} = 0$ for at least one $p \in \{ 1,\ldots,
M\}$ with non-zero $n_p$. It is easy to see that in order to make an SPR
detector work, we need as many different matrix elements $\alpha_{ij}$
as possible. We can therefore choose $n_i=1$ for $i \in \{
2,\ldots,N\}$, where now $N<M$. 

Subsequently, consider the amplitude of $\hat{b}_1^N \hat{b}_l$ for any
$l\in\{ 2,\ldots,M\}$. This stray photon in mode $b_l$ can either originate 
from the input mode or from one of the auxiliary modes, and the coefficient 
is given by 
\begin{equation}
  \alpha_{1l} \cdot \alpha_{11} \cdots \alpha_{N1} + \alpha_{11} \cdot
  \sum_{j=1}^N \alpha_{11} \cdots \alpha_{j-1,1} \alpha_{jl} \cdots
  \alpha_{N1}\; .
\end{equation}
This leads to 
\begin{equation}
  \alpha_{11} \cdot \sum_{j=1}^N (1+\delta_{j1})\; \alpha_{11} \cdots
  \alpha_{j-1,1} \alpha_{jl} \cdots \alpha_{N1} = 0\; . 
\end{equation}
If $\alpha_{p1}=0$, then only one term remains:
\begin{equation}
  \alpha_{11} \cdots \alpha_{pl} \cdots \alpha_{N1} = 0\; , 
\end{equation}
and this forces a second matrix element $\alpha_{q1}$ or $\alpha_{pl}$
to zero. Since this holds true for all $l$, we only have to consider
the case where $\exists\, q:\alpha_{q1}=0$, or $\forall\, l:\alpha_{pl}=0$. 
After all, if only some $\alpha_{pl}$ are zero, there must be a 
$\alpha_{q1}=0$, and there is therefore no need for $\alpha_{pl}=0$. However, 
setting all $\alpha_{pl}$ zero is equivalent to removing the $p^{\rm th}$ 
input photon. Since we set this proof up for an arbitrary number of
auxiliary photons, we only have to consider $\alpha_{q1}=0$. 

The next term in Eq.~(\ref{2double}) the coefficient of which has to
be zero, is $\hat{b}_1^{N-1} \hat{b}_n \hat{b}_{n'}$. With
$\alpha_{p1} = 0$ and $\alpha_{q1} = 0$, the remaining term is
\begin{equation}
  \alpha_{11} \cdots \alpha_{pn} \cdots \alpha_{qn'} \cdots
  \alpha_{N1} = 0\; , 
\end{equation}
and this leads to the conclusion that there must be an $\alpha_{r1} =
0$. 

We can repeat this process such that for each $k$ in the terms
involving $\hat{b}_1^k$, we find an $\alpha_{r1}=0$. There are $N+1$
terms with an overall factor $\hat{b}_1^k$, but there are only $N$
terms $\alpha_{r1}$. We thus have forced all $\alpha_{r1}$ to zero,
and the single-photon input will also have a zero amplitude in mode
$b_1$: 
\begin{equation}\label{Nsingle}
  \hat{a}_1 \prod_{m=2}^N \hat{a}_m ~\rightarrow~ \prod_{m=1}^N
  \left( \sum_{j=1}^M \left. \alpha_{mj} \hat{b}_j
  \right|_{\alpha_{m1}=0} \right) .  
\end{equation}
Since I explicitly constructed the SPR detection device to signal the
presence of a single photon in mode $b_1$, Eq.~(\ref{Nsingle})
contradicts the premise. Hence, an SPR detector of type I is impossible.

Secondly, I consider the type II SPR detector, that is, for at least
one detector, the output is a click when two photons enter the device,
and no click when only one photon enters the device. The impossibility
proof runs along the same lines as the previous proof. With a single
photon in the input mode, the output before the detectors is then
\begin{equation}
  \hat{a}_1 \prod_{m=2}^N \hat{a}_m ~\rightarrow~ \prod_{m=1}^N
  \left( \sum_{j=1}^M \alpha_{mj} \hat{b}_j \right) . 
\end{equation}
Here, I have already used the fact that all $N-1$ auxiliary photons should be
distributed over different modes. Following the same line of reasoning, all 
the $N$ matrix elements $\alpha_{r1}$ with $r\in\{ 1,\ldots,N\}$ are forced 
to zero by the $N$ different $k$-terms proportional to $\hat{b}_1^k$. Again, 
this excludes the possibility for two input photons to trigger the detector
in output mode 1. Therefore, an SPR detector of type II is impossible as
well, which means that single-photon--resolution detection devices with
auxiliary photon-number input states, linear optics and projective
measurements are impossible.

As I mentioned before, a detector cascade can succeed with arbitrary large 
probability of success where the above SPR device fails. Braunstein and I 
have shown in Ref.~\cite{kok01} that such cascades need a large number of 
output modes, and that they are very sensitive to detector losses. 
Notwithstanding these difficulties, Haderka {\em et al}.\ have made 
significant experimental progress towards a working detector cascade. 
Nevertheless, I believe this theorem is an interesting property of 
linear optics with the projective measurements I presented here, and it might 
be useful in guiding our intuition when we try to develop interferometers 
that are to perform special tasks in linear optical quantum computing and 
communication. 

\bigskip

This work was carried out at the Jet Propulsion Laboratory, California
Institute of Technology, under a contract with the National Aeronautics 
and Space Administration. I would like to thank Bill Munro for stimulating 
discussions and valuable comments, and Jonathan Dowling for carefully 
reading this manuscript. I also acknowledge the United States National
Research Council. Part of this research was carried out in the Centre for 
Quantum Computer Technology at the University of Queensland, Australia.

\end{document}